\documentclass[a4paper,12pt]{article}

\usepackage[utf8x]{inputenc}
\usepackage[T1]{fontenc}
\usepackage[colorlinks=true,linkcolor=black,citecolor=black,urlcolor=blue]{hyperref}
\usepackage{geometry}
\usepackage{titlesec}
\titleformat{\section}
{\bfseries\uppercase}{\thesection.}{1em}{}
\titleformat{\subsection}
{\bfseries}{\thesection.\thesubsection.}{1em}{}

\usepackage{graphicx, float, subfig, wrapfig} % used to insert the figure
\usepackage{multirow} % used for the table
\usepackage{cite}
\usepackage{breakurl}
\usepackage{indentfirst}
\usepackage{amsmath, amssymb, amsfonts, bm,stmaryrd}
\usepackage{txfonts}
\usepackage{fourier}
\usepackage{enumitem}
\usepackage[x11names, dvipsnames, svgnames]{xcolor}
\usepackage{enumitem}

\usepackage{tikz, verbatim, pgf, pgfplots, tikz-3dplot,  transparent} % pour superposer images
\usetikzlibrary{arrows, arrows.meta, bending, shapes,positioning,fit,calc, decorations.text, 3d, plotmarks}
\pgfplotsset{compat = newest, ticks=none}

\titlespacing*{\subsection}{0pt}{1.5em}{0.2em}
\titlespacing*{\section}{0pt}{1.5em}{0.2em}

\renewcommand\eqref[1]{Equation~\ref{#1}}

\renewcommand{\thesection}{\arabic{section}}
\renewcommand{\thesubsection}{\arabic{subsection}}

\makeatletter
\renewcommand\@biblabel[1]{#1.}
\makeatother

\newlength{\bibitemsep}
\newlength{\bibparskip}
\let\oldthebibliography\thebibliography
\renewcommand\thebibliography[1]{%
  \oldthebibliography{#1}%
  \setlength{\parskip}{\bibitemsep}%
  \setlength{\itemsep}{\bibparskip}%
}

\newcommand{\YearConf}{2024}

\newcommand{\LogoConf}{logo_IN24.jpg}
\newcommand{\CopyrightConf}{Permission is granted for the reproduction of a fractional part of this paper published in the Proceedings of INTER-NOISE \YearConf ~ \underline{provided permission is obtained} from the author(s) \underline{and credit is given} to the author(s) and these proceedings.}
\usepackage{fancyhdr}
  \fancypagestyle{firststyle}
{   \fancyhf{}
   \fancyfoot[C]{\scriptsize \CopyrightConf}
}
\begin{document}
\thispagestyle{firststyle}

\begin{center}
	\includegraphics[width=2in]{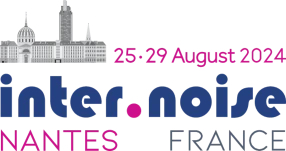}
\end{center}
\vskip.5cm

\begin{flushleft}
\fontsize{16}{20}\selectfont\bfseries Methodology for 3D sound synthesis of directional acoustic sources by higher-order ambisonics
\end{flushleft}
\vskip1cm

\renewcommand\baselinestretch{1}
\begin{flushleft}

Philippe THORNER\footnote{philippe.thorner@lecnam.net}, Éric BAVU, %\footnote{eric.bavu@lecnam.net},
Jean-Baptiste DOC, %\footnote{jean-baptiste.doc@lecnam.net},
Christophe LANGRENNE \\ %\footnote{christophe.langrenne@lecnam.net} \\
Laboratoire de Mécanique des Structures et des Systèmes Couplés du CNAM \\%Conservatoire National des Arts et Métiers \\
2 rue Conté, 75003 Paris, France \\

%\vskip.5cm
%Éric BAVU\footnote{eric.bavu@lecnam.net} \\
%Laboratoire de Mécanique des Structures et des Systèmes Couplés du CNAM \\%Conservatoire National des Arts et Métiers \\
%2 rue Conté, 75003 Paris, France \\
%
%\vskip.5cm
%Jean-Baptiste DOC\footnote{jean-baptiste.doc@lecnam.net} \\
%Laboratoire de Mécanique des Structures et des Systèmes Couplés du CNAM \\%Conservatoire National des Arts et Métiers \\
%2 rue Conté, 75003 Paris, France \\
%
%\vskip.5cm
%Christophe LANGRENNE\footnote{christophe.langrenne@lecnam.net} \\
%Laboratoire de Mécanique des Structures et des Systèmes Couplés du CNAM \\%Conservatoire National des Arts et Métiers \\
%2 rue Conté, 75003 Paris, France \\

%==============================================================================
\end{flushleft}
\textbf{\centerline{ABSTRACT}}\\
\textit{This paper presents the 3D soundfield synthesis of the pressure field radiated by directional acoustic sources using both the multimodal method and higher-order ambisonics (HOA). Ambisonics is a technique for encoding and reproducing measured or modeled (virtual) sound pressure field, based on a decomposition of the acoustic field over spherical harmonics. The directional source considered in this work is an acoustic horn excited by a flat piston. The free-field radiation from this horn is first modeled accurately over a wide frequency range using the multimodal method, which requires relatively low computational resources. This radiated pressure field, collected on a dual-layer sphere of virtual sensors distributed over a Lebedev geometry, allows its projection into the ambisonic domain. The pressure field is then synthesized in the laboratory's 3D 5\textsuperscript{th} order HOA spatialization sphere, which consists of fifty-six loudspeakers. This offers the ability of listening to the radiated sound using a higher-order ambisonic synthesis of a 'virtual' source before it is manufactured. To qualitatively evaluate the performance of the proposed procedure, the transfer function of the synthesized horn is measured around the listening point within the spatialization sphere.
}

%==============================================================================
\section{Introduction}
3D Higher-Order Ambisonics (HOA) is a technique for encoding and decoding a soundfield measured or synthesized in three dimensions. It consists of a decomposition of the acoustic soundfield over a truncated orthogonal basis of spherical harmonics. This decomposition results in the estimation (measures) or the simulation (synthesis) of the ambisonic components. In its early days, this method was only applicable to the first order \cite{gerzon1973}. It was later extended to higher orders \cite{bamford1995, daniel}. Recently, a 3D 5\textsuperscript{th} order spatialization sphere, consisting of fifty-six loudspeakers driven by HOA, has been developed \cite{lecomtePhD, lecomte2015real} and has been tested for 3D soundfield synthesis, which has successfully been applied to moving sources localization applications \cite{pujol, pujol2021}.

\begin{figure}[H]
	\centering
	\begin{tikzpicture}
		% Texte sources
		%\node (1) {\includegraphics[width=.1\textwidth]{trombone}};
		% Images
		% Modélisation du pavillon
		\node[node distance=4.cm] (2) {{\transparent{.4} \includegraphics[width=.1\textwidth]{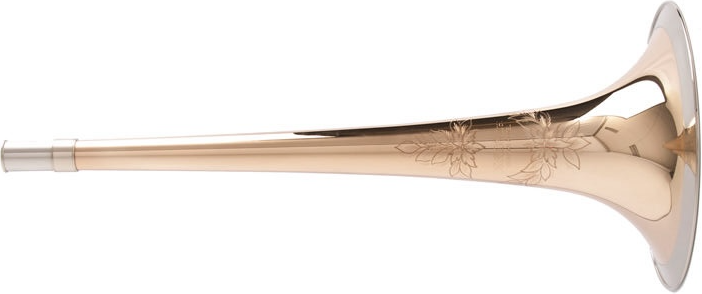}}};
		\node[node distance=0.5cm, below of=2] {\footnotesize $\textcolor{red}{S(x) = f(S_0, x)}$};
		% Flèche multimodale
		%\draw[-stealth, line width=1.mm, black] (1) to node[midway, above] {\textcolor{black}{\textbf{\scriptsize Source}}} node[midway, below] {\textcolor{black}{\textbf{\scriptsize modelling}}} (2);
		
		\node[node distance=0.2cm] [above of=2] {
			\begin{tikzpicture}[scale=.41]
				\draw[line width=.75mm, red] (5.7, 5.05) -- (7.3, 5.1) -- (8, 5.135) -- (8.45, 5.1755);
				\draw[line width=.75mm, red] (8.445, 5.175) .. controls (8.8,5.2) and (9.3,5.25) .. (9.5,5.52);
				\draw[line width=.75mm, red] (9.49,5.504) .. controls (9.55,5.62) and (9.6,5.7) .. (9.61,5.8);
			\end{tikzpicture}
		};
		
		%Cartographie de pression
		\node[node distance=5.cm] (3) [right of =2]{\includegraphics[width=.2\textwidth]{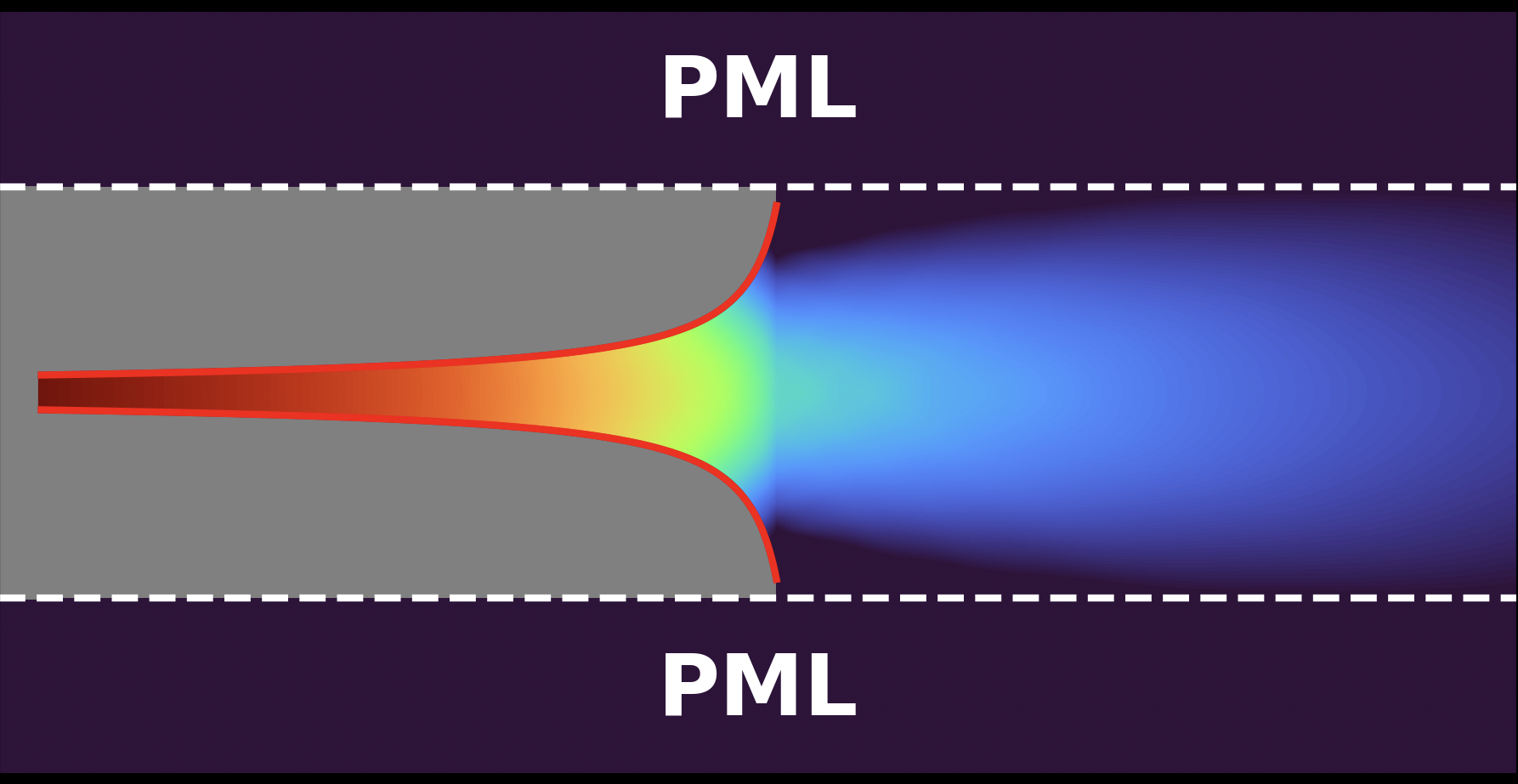}};
		% Flèche simulation
		\draw[-stealth, line width=1.mm, black] (2) to node[left, above] {\textcolor{black}{\textbf{\scriptsize Simulation}}} (3);
		
		% Définition du domaine sur lequel sont calculées les fonctions de transfert
		\tdplotsetmaincoords{90}{0}
		% Tracé du domaine, cercle blanc
		\node[node distance=4.7cm, circle, minimum size=.6cm, fill=white, xshift=1.cm] [right of=2] {};
		
		\node[node distance=4.2cm, xshift=1.5cm] (4) [right of=2] {
			\begin{tikzpicture}[tdplot_main_coords, scale=.72, outer sep=0pt]
				% Tracé des points de Lebedev correspondant au degré 11 (50 points)
				% Points visibles aux premier plan
				% Projection sur la sphère des sommets d’un octaèdre
				\filldraw[black] (6.5, 0, 2.2) circle (.25pt);
				\filldraw[black] (6.7, 0, 2) circle (.25pt);
				\filldraw[red] (6.5, .2, 2) circle (.25pt);
				\filldraw[black] (6.3, 0, 2) circle (.25pt);
				\filldraw[black] (6.5, 0, 1.8) circle (.25pt);
				% Projection sur la sphère des arrêtes d’un octaèdre
				\filldraw[black] (6.21716, 0, 2.28284) circle (.25pt);
				\filldraw[black] (6.21716, 0.28284, 2) circle (.25pt);
				\filldraw[black] (6.78284, 0, 1.7172) circle (.25pt);
				\filldraw[black] (6.5, 0.28284, 1.7172) circle (.25pt);
				\filldraw[black] (6.21716, 0, 1.7172) circle (.25pt);
				\filldraw[black] (6.78284, 0, 2.28284) circle (.25pt);
				\filldraw[black] (6.5, 0.28284, 2.28284) circle (.25pt);
				\filldraw[black] (6.78284, 0.28284, 2) circle (.25pt);
				% projections sur la sphère du centre des faces d’un octaèdre
				\filldraw[black] (6.2691, 0.2309, 2.2309) circle (.25pt);
				\filldraw[black] (6.2691, 0.2309, 1.76906) circle (.25pt);
				\filldraw[black] (6.7309, 0.2309, 2.2309) circle (.25pt);
				\filldraw[black] (6.7309, 0.2309, 1.76906) circle (.25pt);
				% projections de points situés sur la bissectrice du triangle équilatéral formé par la face d’un octaèdre
				\filldraw[black] (6.3794, 0.3618, 2.1206) circle (.25pt);
				\filldraw[black] (6.1382, 0.1206, 2.1206) circle (.25pt);
				\filldraw[black] (6.3794, 0.3618, 1.8794) circle (.25pt);
				\filldraw[black] (6.1382, 0.1206, 1.8794) circle (.25pt);
				\filldraw[black] (6.6206, 0.1206, 1.6382) circle (.25pt);
				\filldraw[black]  (6.3794, 0.1206, 1.6382) circle (.25pt);
				\filldraw[black] (6.6206, 0.1206, 2.3618) circle (.25pt);
				\filldraw[black] (6.3794, 0.1206, 2.3618) circle (.25pt);
				\filldraw[black] (6.8618, 0.1206, 2.1206) circle (.25pt);
				\filldraw[black] (6.6206, 0.3618, 2.1206) circle (.25pt);
				\filldraw[black] (6.8618, 0.1206, 1.8794) circle (.25pt);
				\filldraw[black] (6.6206, 0.3618, 1.8794) circle (.25pt);
			\end{tikzpicture}
		};
		
		% Fonction de transfert
		\node[node distance=1.9cm] (5) [below of=3] {\includegraphics[width=.2\textwidth]{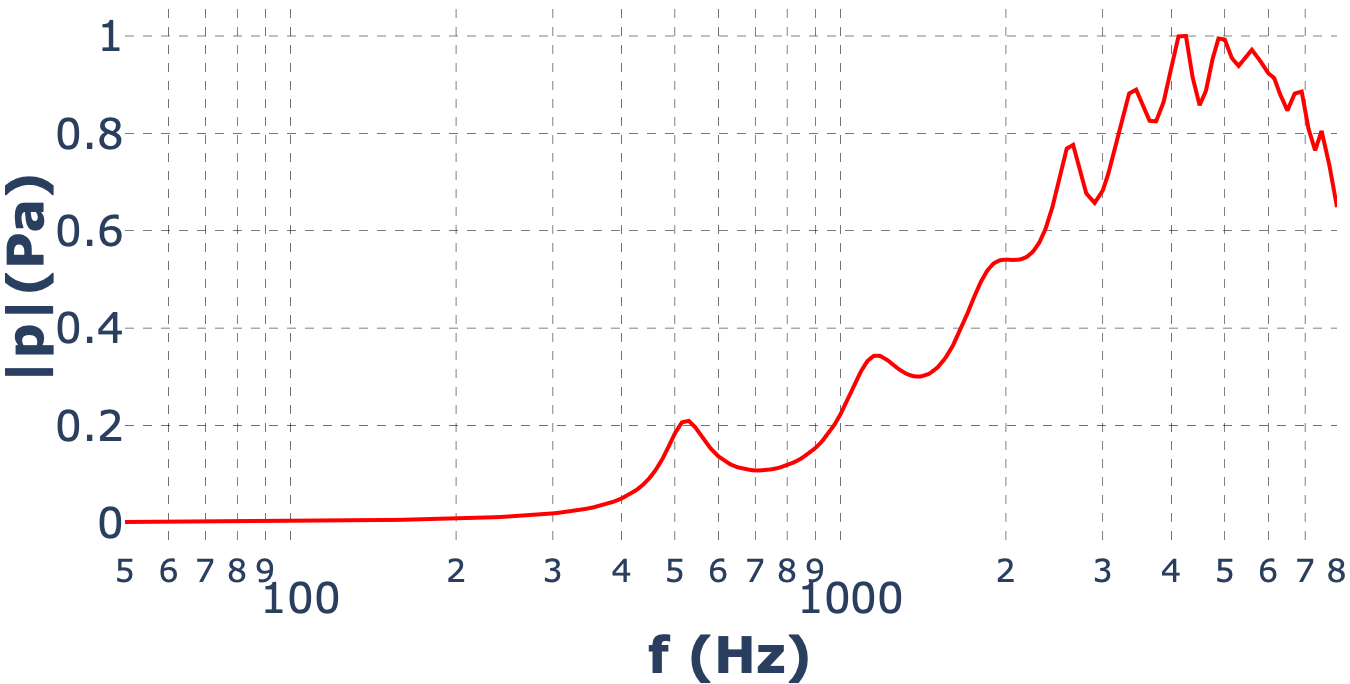}};
		\node[node distance=1.2cm] (6) [below of=3] {};
		%\node (7.65, -1.9) {\footnotesize Exemple d'une fonction de transfert};
		% flèche FT
		\draw[-stealth, align=center, red] (4.center) to (6);

		% Encodage ambisonique 
		\node[node distance=4.5cm] (7) [right of=3] {};
		\node[node distance=6.5cm, yshift=-1.cm] (8) [right of=3] {
			% Synthèse sonore dans la sphère
			\begin{tikzpicture}[scale=.7]
				\begin{scope}[shift={(0.8cm, 1.2cm)}, scale=2.6]
					\tdplotsetmaincoords{60}{110}
					
					\tdplotsetthetaplanecoords{45}
					\tdplotdrawarc[tdplot_rotated_coords, thin, gray]{(0,0,0)}{1}{150} {-20}{anchor=west}{}
					\tdplotdrawarc[tdplot_rotated_coords, thin, gray, dashed]{(0,0,0)}{1}{0} {-210}{anchor=west}{}
					% Partie de l'octaèdre
					\tdplotdrawarc[tdplot_rotated_coords, thin, gray]{(0,0,0)}{1}{90} {0}{anchor=west}{}
					\tdplotsetthetaplanecoords{-45}
					\tdplotdrawarc[tdplot_rotated_coords, thin, gray]{(0,0,0)}{1}{125} {-55}{anchor=west}{}
					\tdplotdrawarc[tdplot_rotated_coords,  thin, gray, dashed]{(0,0,0)}{1}{0} {-235}{anchor=west}{}
					\tdplotsetthetaplanecoords{-180}
					\tdplotdrawarc[tdplot_rotated_coords, thin, gray]{(0,0,0)}{1}{35} {-125}{anchor=west}{}
					\tdplotdrawarc[tdplot_rotated_coords, thin, gray, dashed]{(0,0,0)}{1}{-120} {-325}{anchor=west}{}
					% Partie de l'octaèdre
					\tdplotdrawarc[tdplot_rotated_coords, thin, gray]{(0,0,0)}{1}{0} {-90}{anchor=west}{}
					\tdplotsetthetaplanecoords{-90}
					\tdplotdrawarc[tdplot_rotated_coords, thin, gray]{(0,0,0)}{1}{45} {-135}{anchor=west}{}
					\tdplotdrawarc[tdplot_rotated_coords, thin, gray, dashed]{(0,0,0)}{1}{0} {-315}{anchor=west}{}
					% Partie de l'octaèdre
					\tdplotdrawarc[tdplot_rotated_coords, thin, gray]{(0,0,0)}{1}{0} {-90}{anchor=west}{}
					
					\begin{scope}[tdplot_rotated_coords, canvas is yz plane at x=0]
						\draw[thin, gray, dashed] (-1,0) arc(-180:45:1);
						\draw[thin, gray] (-1,0) arc(180:20:1);
						% Partie de l'octaèdre
						\draw[thin, gray] (-1,0) arc(180:90:1);
					\end{scope}
					
					\tdplotsetrotatedcoords{-20}{68.5}{0}
					\begin{scope}[tdplot_rotated_coords, canvas is yz plane at x=0]
						\draw[thin, gray] (-1,0) arc(-180:-20:1);
						\draw[thin, gray] (-1,0) arc(-180:-210:1);
						\draw[thin, gray, dashed] (-1,0) arc(180:-20:1);
						% Partie de l'octaèdre
						\draw[thin, gray] (-.8,-0.6) arc(-143:-53:1);
					\end{scope}
					
					\tdplotsetrotatedcoords{-21.8}{8.5}{0}
					\begin{scope}[tdplot_rotated_coords, canvas is yz plane at x=0]
						\draw[thin, gray] (-1,0) arc(-180:-35:1);
						\draw[thin, gray, dashed] (-1,0) arc(180:-35:1);
						\draw[thin, gray] (-1,0) arc(-180:-215:1);
					\end{scope}
					
					\tdplotsetrotatedcoords{90}{45}{0}
					\begin{scope}[tdplot_rotated_coords, canvas is yz plane at x=0]
						\draw[thin, gray, dashed] (-1,0) arc (180:-120:1);
						\draw[thin, gray] (-1,0) arc (180:45:1);
						\draw[thin, gray] (-1,0) arc (180:230:1);
						% Partie de l'octaèdre
						\draw[thin, gray] (-1,0) arc (180:90:1);
					\end{scope}
					
					\tdplotsetrotatedcoords{0}{45}{0}
					\begin{scope}[tdplot_rotated_coords, canvas is xy plane at z=0]
						\draw[thin, gray, dashed] (0,1) arc (90:-215:1);
						\draw[thin, gray] (0,1) arc (90:30:1);
						\draw[thin, gray] (0,1) arc (90:145:1);
					\end{scope}
					
					\tdplotsetrotatedcoords{0}{0}{0}
					\begin{scope}[tdplot_rotated_coords]		
						% Tracé des points de Lebedev correspondant au degré 11 (50 points)
						% Points non-visibles aux premier plan
						% Projection sur la sphère des sommets d’un octaèdre
						\filldraw[color=black!60] (-1, 0, 0) circle (.75pt) node[above right] {};
						\filldraw[color=black!60] (0, -1, 0) circle (.75pt) node[below right] {};
						\filldraw[color=black!60] (0, 0, -1) circle (.75pt) node[above right] {};
						% Projection sur la sphère des arrêtes d’un octaèdre
						\filldraw[blue] (-0.70711, 0, 0.70711) circle (.75pt) node[above right] {};
						\filldraw[blue] (0, -0.70711, 0.70711) circle (.75pt) node[above right] {};
						\filldraw[blue] (-0.70711, 0.70711, 0) circle (.75pt) node[below right] {};
						\filldraw[blue] (-0.70711, -0.70711, 0) circle (.75pt) node[above right] {};
						\filldraw[blue] (0.70711, 0, -0.70711) circle (.75pt) node[above right] {};
						\filldraw[blue] (0, 0.70711, -0.70711) circle (.75pt) node[above right] {};
						\filldraw[blue] (-0.70711, 0, -0.70711) circle (.75pt) node[above right] {};
						\filldraw[blue] (0, -0.70711, -0.70711) circle (.75pt) node[above right] {};
						% projections sur la sphère du centre des faces d’un octaèdre
						\filldraw[green] (-0.57735, 0.57735, 0.57735) circle (.75pt) node[above right] {};
						\filldraw[green] (-0.57735, -0.57735, 0.57735) circle (.75pt) node[above right] {};
						\filldraw[green] (-0.57735, 0.57735, -0.57735) circle (.75pt) node[above right] {};
						\filldraw[green] (-0.57735, -0.57735, -0.57735) circle (.75pt) node[above right] {};
						\filldraw[green] (0.57735, -0.57735, -0.57735) circle (.75pt) node[above] {};
						% projections de points situés sur la bissectrice du triangle équilatéral formé par la face d’un octaèdre
						\filldraw[red] (-0.3015, -0.3015, 0.9045) circle (.75pt) node[above right] {};
						\filldraw[red] (-0.3015, 0.9045, 0.3015) circle (.75pt) node[above right] {};
						\filldraw[red] (-0.9045, 0.3015, 0.3015) circle (.75pt) node[above left] {};
						\filldraw[red] (-0.9045, -0.3015, 0.3015) circle (.75pt) node[above right] {};
						\filldraw[red] (-0.3015, -0.9045, 0.3015) circle (.75pt) node[above right] {};
						\filldraw[red] (0.3015, -0.9045, 0.3015) circle (.75pt) node[above right] {};
						\filldraw[red] (-0.3015, 0.9045, -0.3015) circle (.75pt) node[above right] {};
						\filldraw[red] (-0.9045, 0.3015, -0.3015) circle (.75pt) node[above right] {};
						\filldraw[red] (-0.9045, -0.3015, -0.3015) circle (.75pt) node[above right] {};
						\filldraw[red] (-0.3015, -0.9045, -0.3015) circle (.75pt) node[above right] {};
						\filldraw[red] (0.3015, -0.9045, -0.3015) circle (.75pt) node[below right] {};
						\filldraw[red] (0.3015, 0.3015, -0.9045) circle (.75pt) node[above] {};
						\filldraw[red]  (-0.3015, 0.3015, -0.9045) circle (.75pt) node[above right] {};
						\filldraw[red] (-0.3015, -0.3015, -0.9045) circle (.75pt) node[above right] {};
						\filldraw[red] (0.3015, -0.3015, -0.9045) circle (.75pt) node[above right] {};
						
					\end{scope}
					
					% Tracé d'une sphère transparente pour donné un rendu 3D
					%\fill[ball color=lightgray, opacity = 0.2] (0,0) circle (1);
					
					% Points non-visibles aux premier plan
					\begin{scope}[tdplot_rotated_coords]
						% Projection sur la sphère des sommets d’un octaèdre
						\filldraw[black] (0, 0, 1) circle (.75pt) node[above right] {};
						\filldraw[black] (1, 0, 0) circle (.75pt) node[above right] {};
						\filldraw[black] (0, 1, 0) circle (.75pt) node[above right] {};
						% Projection sur la sphère des arrêtes d’un octaèdre
						\filldraw[blue] (0.70711, 0, 0.70711) circle (.75pt) node[above right] {};
						\filldraw[blue] (0, 0.70711, 0.70711) circle (.75pt) node[above right] {};
						\filldraw[blue] (0.70711, 0.70711, 0) circle (.75pt) node[above right] {};
						\filldraw[blue] (0.70711, -0.70711, 0) circle (.75pt) node[above right] {};
						% projections sur la sphère du centre des faces d’un octaèdre
						\filldraw[green] (0.57735, 0.57735, 0.57735) circle (.75pt) node[above right] {};
						\filldraw[green] (0.57735, -0.57735, 0.57735) circle (.75pt) node[above right] {};
						\filldraw[green] (0.57735, 0.57735, -0.57735) circle (.75pt) node[above right] {};
						% projections de points situés sur la bissectrice du triangle équilatéral formé par la face d’un octaèdre
						\filldraw[red] (0.3015, 0.3015, 0.9045) circle (.75pt) node[above right] {};
						\filldraw[red] (-0.3015, 0.3015, 0.9045) circle (.75pt) node[above right] {};
						\filldraw[red] (0.3015, -0.3015, 0.9045) circle (.75pt) node[above right] {};
						\filldraw[red] (0.9045, 0.3015, 0.3015) circle (.75pt) node[above right] {};
						\filldraw[red] (0.3015, 0.9045, 0.3015) circle (.75pt) node[above right] {};
						\filldraw[red] (0.9045, -0.3015, 0.3015) circle (.75pt) node[above right] {};
						\filldraw[red] (0.9045, 0.3015, -0.3015) circle (.75pt) node[above right] {};
						\filldraw[red] (0.3015, 0.9045, -0.3015) circle (.75pt) node[above right] {};
						\filldraw[red] (0.9045, -0.3015, -0.3015) circle (.75pt) node[above right] {};
					\end{scope}
					
					% tête
					\node at (-0.01, 0.01) {\includegraphics[width=.02\textwidth]{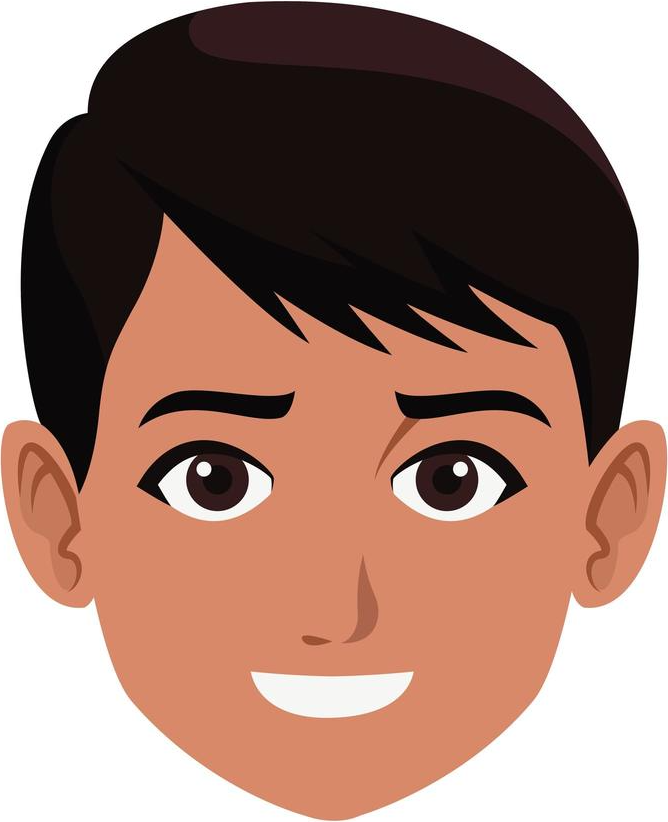}};
				\end{scope}
				
				\node at (.8, 4.1) {\scriptsize \textbf{3D Sound Synthesis}};
			\end{tikzpicture}
		};
		
		% Flèche simulation
		\draw[-stealth, line width=1.5mm, black] (3) to node[midway, above] {\textcolor{black}{\textbf{\scriptsize Ambisonic}}} node[midway, below] {\textcolor{black}{\textbf{\scriptsize encoding}}} (7);
		
	\end{tikzpicture}
	
	%\vspace{-.5cm}
	\caption{3D sound synthesis of a horn by HOA}
	\label{methodo}
\end{figure}
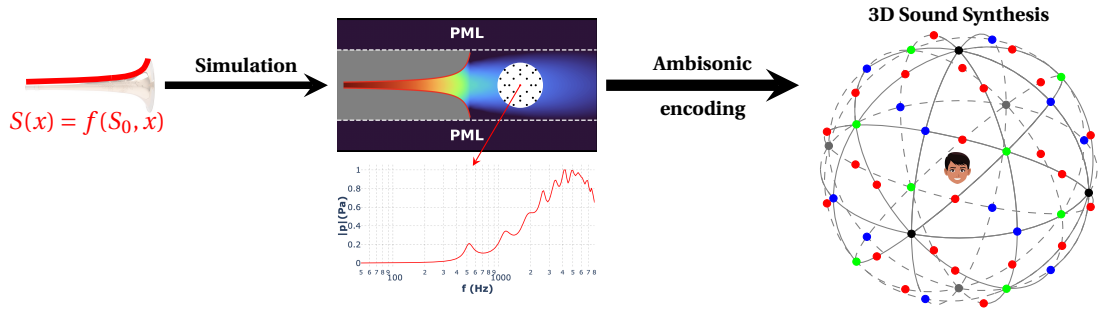

The acoustic source studied in this paper is an acoustic horn, whose radiation is frequency-dependent. The 3D synthesis of the source soundfield is illustrated in Figure \ref{methodo}. As a first step, the pressure field radiated by the horn has to be simulated over a computational domain consisting of a 100 points distributed over a dual-layer Lebedev geometry. These points are then projected into the ambisonic domain.
As a final step, the pressure field is synthesized in the fifty-six-loudspeaker sphere. This allows the source to be measured or listened to in the sphere prior to being manufactured.
The simulation methodology used to reproduce the radiated pressure field needs to operate over a wide range of frequencies and offer low computational costs. This is the reason why the multimodal formulation was used. This formulation is a semi-analytical method taking benefits from the preferential direction of guided propagation \cite{pagneux1996}. This method is commonly used to study the propagation of sound waves in waveguides. In particular, Guennoc et al. \cite{doc2} developed a general formulation to calculate the pressure field in waveguides with a variable section, such as horns. Félix et al. \cite{doc1} introduced the use of a Perfectly Matched Layer (PML) in the multimodal formulation to simulate the radiation of a waveguide in free field conditions.

\bigskip
In the first section of this paper, the computation of the pressure field using the multimodal formulation is outlined. A second section describes the ambisonic encoding of the simulated pressure field and its subsequent synthesis in the spatialization sphere. In the third section, results are presented and analyzed. The two last sections are dedicated to conclusions and future developments.

%==============================================================================
\section{Computation of the radiated pressure field}
\begin{figure}[H]
	\centering
	\begin{tikzpicture}[scale=.73]
		% PML
		\draw[black, line width=7pt] (-4,7) -- (11,7); 				% paroi supérieure externe du guide B
		\fill [yellow!20!white] (-4,5.8) rectangle (11,7); 	   % remplissage de la couche PML supérieure
		\node[text=DarkBlue] at (3,6.4) {\large \textbf{PML}};     % Texte PML en haut en gris dans la couche en jaune
		\node at (7.5,4.5) {\textbf{\small WGB}};
		
		% Tracé du guide annulaire A
		\draw[dashed] (0,5.08) -- (4.8,5.08); 
		\draw[dashed] (0,4.92) -- (5.5,4.92);
		\draw[gray,thick] (0,5.08) -- (4.015,5.08); 						 % paroi externe du guide A
		\draw[gray,thick] (0,4.92) -- (4.015,4.92); 			       % paroi interne du rectangle au-dessus du pavilon, guide A
		\draw[gray,thick] (4,4.92) -- (4,5.08); 								 % paroi frontale du pavillon
		\node at (2,5.4) {\textbf{\small WGC}};
		
		% Aire du guide contenant le pavillon
		\shade[top color=lightgray,bottom color=lightgray] (0,3.7) parabola (4,4.92) -| (4,3.7);
		\shade[top color=lightgray,bottom color=lightgray] (0,3.2) rectangle (4,3.72) -| (0,3.72);
		\shade[top color=gray,bottom color=gray] (0,3.7) parabola (4,4.92) -| (0,4.92);
		\shade[top color=gray,bottom color=gray] (0,4.92) rectangle (4,5.08) -| (0,5.08);
		% Tracé du pavillon
		\draw[gray, thick] (0,3.7) parabola (4,4.92);
		\draw[gray, line width=2mm] (.125,3.2) -- (.125,3.7);
		\node at (2.4,3.6) {\textbf{\small WGA}};
		
		% Tracé des axes
		%axis
		\draw[thick, dashed, -stealth] (-4,3.2) -- (11,3.2) node[anchor=north east] {\small $x$};
		\draw[thick, dashed, -stealth] (4,3.2) -- (4,7.7) node[anchor=south east] {\small $r$};
		\node at (4,2.9) {\textbf{\small $0$}};
		
		% Cotes
		\draw[stealth-stealth] (4.5,3.2) -- (4.5,4.92);
		\node at (4.8,3.95) {\small a};
		\draw[stealth reversed-stealth reversed] (4.7,4.81) -- (4.7,5.19);
		\node at (5.,5) {\small e};
		\draw[stealth - stealth] (4.3,4.92) -- (4.3,5.8);
		\node at (4.6,5.36) {\small $\epsilon$};
		\draw[stealth - stealth] (10,3.2) -- (10,7);
		\node at (10.3,5) {\small b};
		\draw[stealth - stealth] (9,5.8) -- (9,7);
		\node at (9.3,6.4) {\small $\delta$};
		
	\end{tikzpicture}
	\caption{Radiation in free field of a baffled horn using the multimodal formulation with a PML}
	\label{config}
\end{figure}
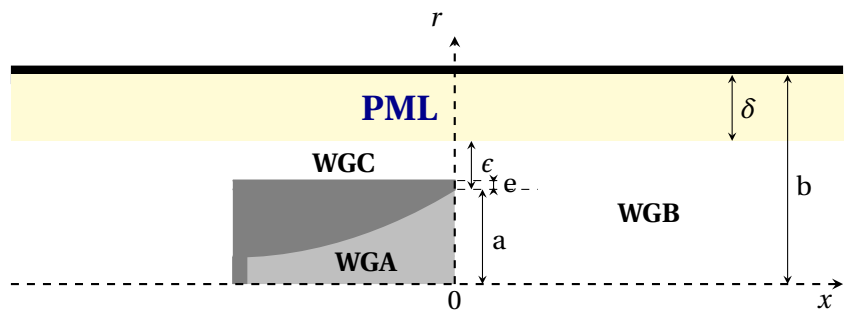

Figure \ref{config} illustrates the configuration for the study of propagation in the horn and its radiation using the multimodal formulation with a PML.
The studied baffled Wave Guide A (WGA) is placed in a larger Wave Guide B (WGB), with an absorbent layer (PML) of thickness $\delta$ on its walls, in order to simulate its radiation in free field conditions. The Wave Guide C (WGC) ensures that the baffle is taken into account in the radiation condition. The pressures inside the waveguides A, B and C are solutions of the Helmoltz's equation
\begin{equation} \label{waveeq}
	(\Delta + k^2)p = 0,
\end{equation}
and can be written as a modal decomposition
\begin{equation} \label{champPression}
	p_{\eta}(r,x) = \sum_{n=0}^{N_{\eta}} P_{{\eta}_n}(x) \varphi_{{\eta}_n}(r),
\end{equation}
where $\eta = A, B, C$ and $\varphi_{{\eta}_n}$ is the modal function of the $n^{th}$ mode, element of the orthogonal basis built with Bessel functions of the first and second kinds. WGB is assumed to be of infinite length. These assumptions set the radiation condition at the output of WGA. By writing the continuity of pressures and pressure gradients at the junction of the waveguides at $x=0$ the radiation impedance yields:
\begin{equation}
	Z_{A_{out}} = F Z_B \left[I + G^T Y_C G Z_B \right]^{-1} F^T,
\end{equation}
where $F$ and $G$ represent the inter-modal coupling between respectively WGA \& WGB and WGC \& WGB. $Z_B$ is the characteristic impedance of WGB. $Y_C$ is the admittance of WGC brought back to the junction at $x = 0$. $I$ is the identity matrix.

\begin{figure}[H]
	\subfloat[pressure map inside the horn, $ka = 16.7$]{
		\includegraphics[width=.5125\textwidth]{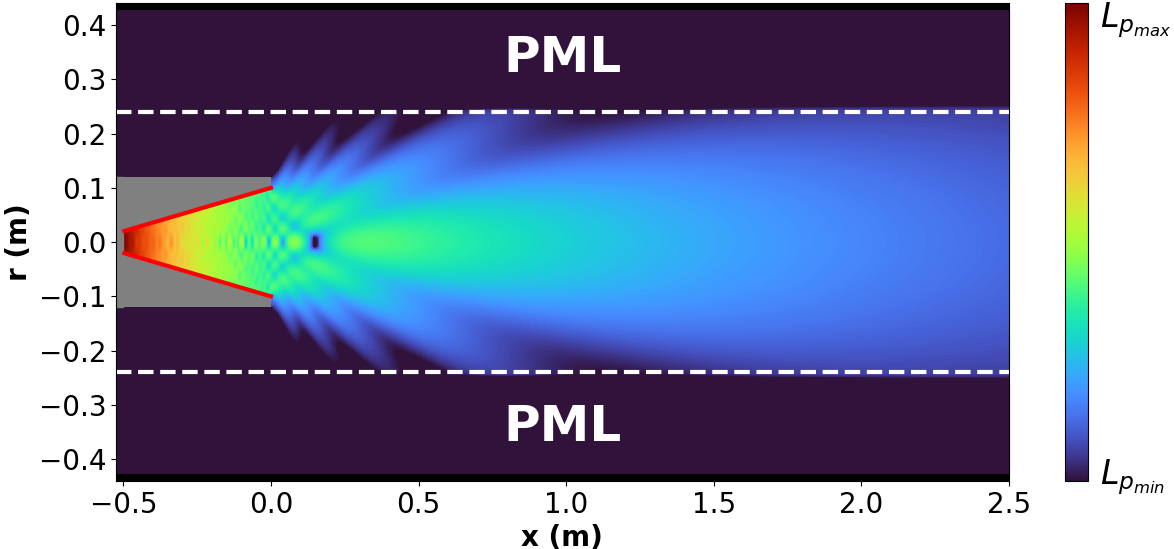} \label{map}
	}
	\subfloat[on-axis frequency response of the horn at 2 m]{
		\includegraphics[width=.4875\textwidth]{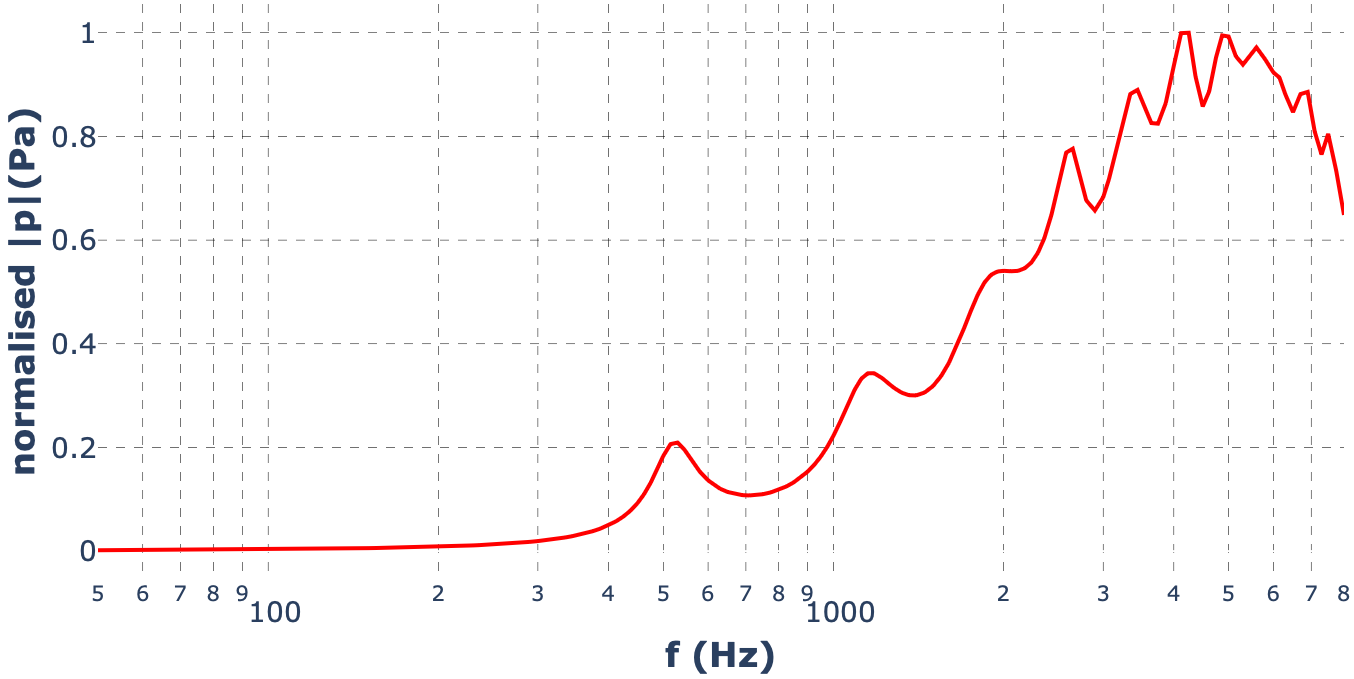} \label{ft}
	}
	\caption{a) Pressure map at $ka = 16.7$ and b) frequency response on-axis at 2 m for a conical horn ($r_{in}=2$ cm, $r_{out}=10$ cm, $L=50$ cm)}
\end{figure}

Equation (\ref{waveeq}) and the pressure gradient $q = \partial_x p$ are then both projected onto the modes to obtain a system of equations relating pressure, pressure gradient and their first order derivatives. Substituting the gradient by $Q = YP$ in this system leads to the Riccati equation for the admittance:
\begin{equation} \label{riccati}
	Y' = -Y^2 + Y F + F^T Y - K^2,
\end{equation}
where $K$ is the dispersion matrix, which characterizes the modes propagation inside the horn. Equation (\ref{riccati}), being non-linear, is numerically solved with the help of a Magnus \cite{magnus1954} scheme. The resolution requires an initial condition, which is known at the interface $x=0$ and is described by $Y_{|_{x=0}}~=~1/Z_{A_{out}}$. As a result of this process, the impedance is now determined at every point inside the horn. An initial condition for the pressure is then imposed to compute the pressure field inside the horn. Using the coupling between the guides, the radiated pressure field outside the horn can now be simulated. It is then possible either to plot a pressure map at a given frequency (see Figure \ref{map}) or a frequency response in free field (see Figure \ref{ft}).

\section{Ambisonic encoding}
First, the pressure field radiated by the horn is computed for 100 virtual sensors distributed on a dual-layer Lebedev geometry using the methodology presented in the previous section. This allows  the coefficients of the spherical Fourier transform (SFT), $P_{nm}$, to be calculated using a discrete version of the following equation:
\begin{equation} \label{sftcoef}
	P_{nm} = \int_0^{2\pi} \int_0^\pi p(\theta,\delta) Y_n^m(\theta_s,\delta_s) \sin(\delta) \text{d}\delta \text{d}\theta,
\end{equation}
where $p(\theta,\delta)$ is the radiated pressure field simulated with the multimodal method and $Y_n^m(\theta,\delta)$ is the normalized spherical harmonic of indices ($m, n$).
Since the pressure field cannot be computed over the continuous surface of a sphere, a Lebedev grid \cite{lecomte2016, lebedev1975} is used to discretize the surface, which ensures a minimization of the projection's error and provides the best performances of soundfield reproduction in the 3D spatialization sphere. After the discretization of the projection surface, equation (\ref{sftcoef}) writes
\begin{equation}
	P_{nm} = \sum_{l=0}^L w_l p(\theta_l,\delta_l) Y_n^m(\theta_l,\delta_l),
\end{equation}
where $w_l$ is the weight of the $l^{th}$ Lebedev's point \cite{lebedev1975}.

Finally, the ambisonic components are computed using the dual layer approach proposed by Williams \cite{williams1999fourier}:
\begin{equation}\label{compoambi}
	B_{nm} = \dfrac{P_{nm}(r_1) h_n^{(2)}(kr_2) - P_{nm}(r_2)  h_n^{(2)}(kr_1)}{i^n \left[j_n(kr_1) h_n^{(2)}(kr_2) - j_n(kr_2) h_n^{(2)}(kr_1) \right]},
\end{equation}
where $h_n^{(2)}$ and $j_n$ are respectively the spherical Hankel function of the second kind and the spherical Bessel function of the first kind.

%==============================================================================
\section{Results}
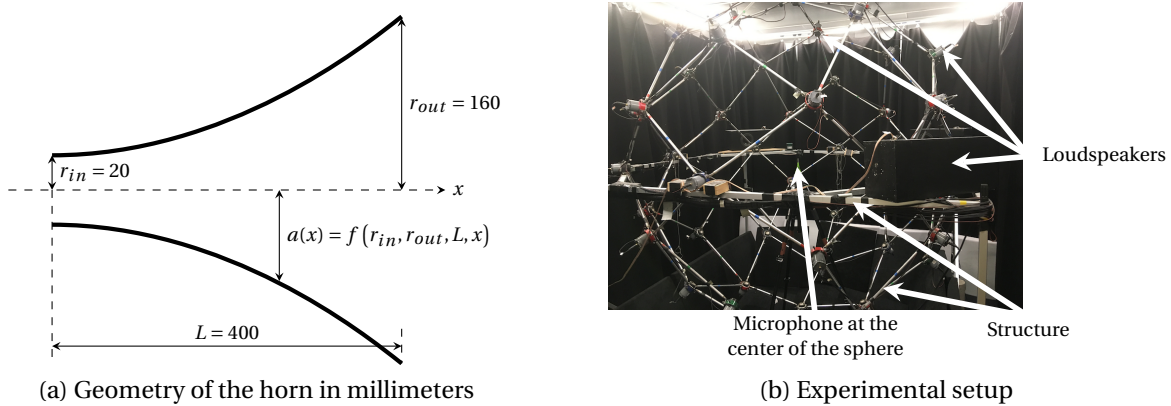
\begin{figure}[H]
	\centering
	\subfloat[Geometry of the horn in millimeters]{\label{horngeometry}
		\scriptsize
		\begin{tikzpicture}[scale=.23]
			% Profil quelconque
			\draw[ultra thick, black] (0, 2) parabola (20, 10);
			\draw[ultra thick, black] (0, -2) parabola (20, -10);
			\draw[black, stealth-stealth] (0, 0) -- (0, 2) node[midway, right] {$r_{in} = 20$};
			\draw[black, stealth-stealth] (20, 0) -- (20, 9.9) node[midway, right] {$r_{out} = 160$};
			\draw[black, stealth-stealth] (13, 0) -- (13, -5.25) node[midway, right] {$a(x) = f \left(r_{in}, r_{out}, L, x \right)$};
			\draw[dashed, black, -stealth] (-2.5, 0) -- (22.5, 0) node[right] {$x$};
			\draw[dashed] (0, 0) -- (0, -10);
			\draw[dashed] (20, -8) -- (20, -10);
			\draw[black, stealth-stealth] (0, -9) -- (20, -9) node[midway, above] {$L = 400$};
		\end{tikzpicture}
	} \quad \quad \quad
	\subfloat[Experimental setup]{\label{sphere}
		\scriptsize
		\begin{tikzpicture}
			\node (1) {\includegraphics[width=.32\textwidth]{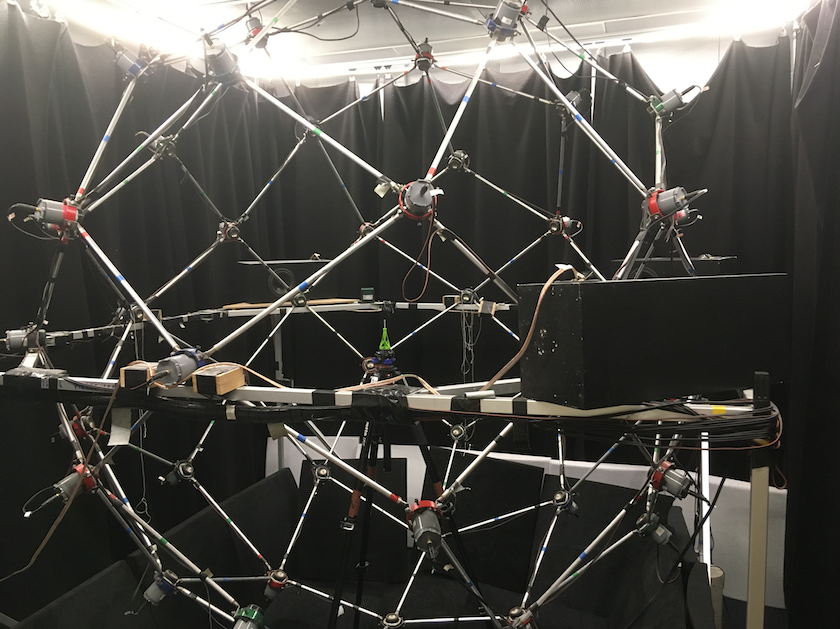}};
			\node[node distance=0.05cm, xshift=-0.25cm] (2) [below of=1] {};
			\draw[stealth-, line width=.75mm, white] (2) to (1.south);
			\node[node distance=2.4cm, align=center] (3) [below of =1] {Microphone at the \\ center of the sphere};
			\node[node distance=1.65cm, text=white] (4) [above of=1] {};
			\node[node distance=1.5cm, xshift=1.6cm, text=white] (5) [above of=1] {};
			\node[node distance=1.7cm, yshift=-.1cm, text=white] (6) [right of=1] {};
			\draw[stealth-, line width=.75mm, white] (4) to (1.east);
			\draw[stealth-, line width=.75mm, white] (5) to (1.east);
			\draw[stealth-, line width=.75mm, white] (6) to (1.east);
			\node[node distance=3.8cm, align=center] (7) [right of =1] {Loudspeakers};
			\node[node distance=.8cm, yshift=-1.73cm, text=white] (8) [right of=1] {};
			\node[node distance=.4cm, yshift=-.55cm, text=white] (9) [right of=1] {};
			\draw[stealth-, line width=.75mm, white] (8) to (1.south east);
			\draw[stealth-, line width=.75mm, white] (9) to (1.south east);
			\node[node distance=2.3cm, xshift=2.8cm, align=center] (10) [below of =1] {Structure};
			
		\end{tikzpicture}
	}
	\caption{a) Horn's geometry and b) experimental setup for measurements}
	\label{mesure}
\end{figure}

\begin{figure}[H]
	\centering
	\includegraphics[width=.7\textwidth]{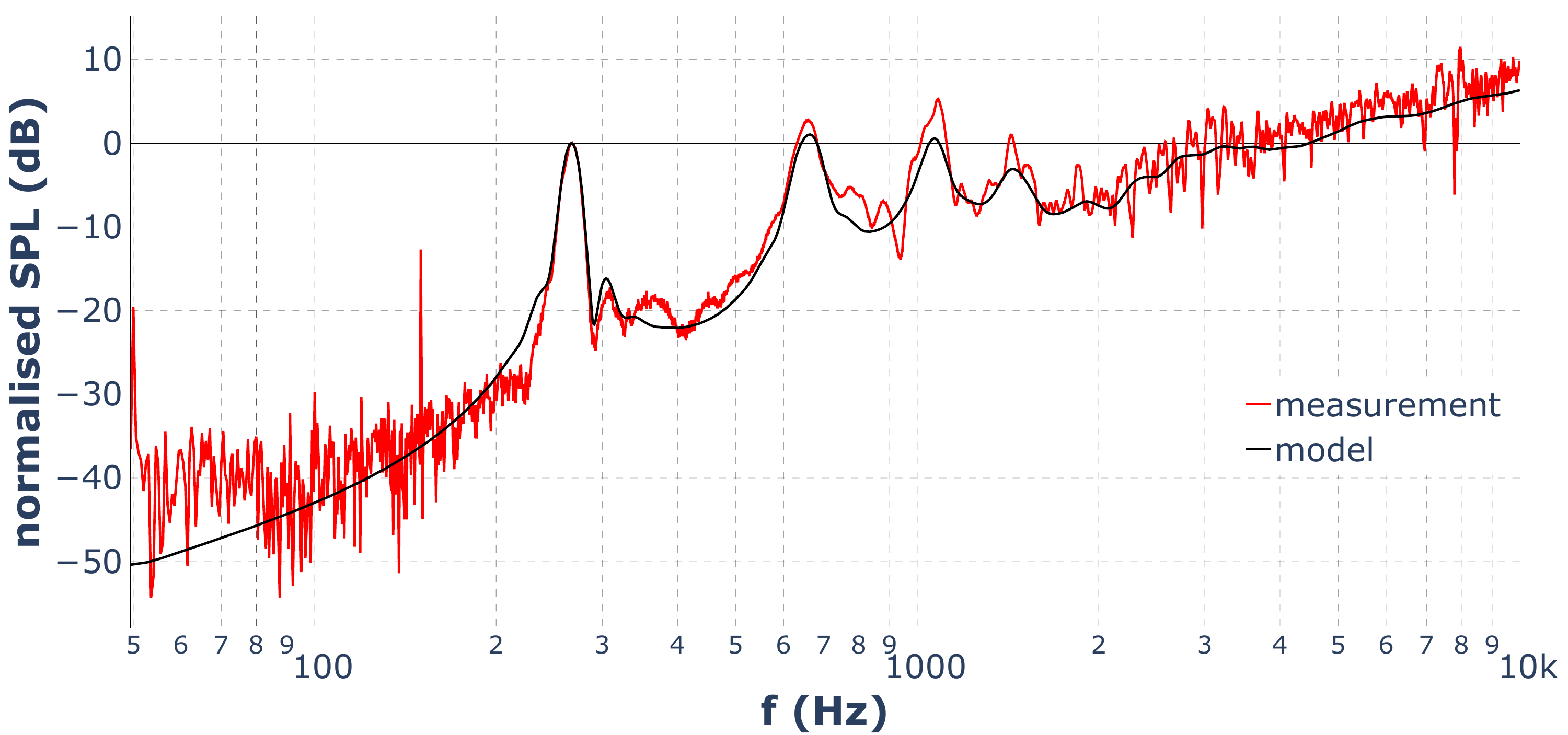}
	\caption{On-axis Transfer Function of a Bessel horn at 2 m ($r_{in}=2$ cm, $r_{out}=16$ cm, $L=40$ cm)}
	\label{result}
\end{figure}

The proposed methodology has been applied to a Bessel horn placed on-axis at 2 meters from the listening point. The horn geometry is illustrated in Figure \ref{horngeometry}. The transfer function of the horn's virtual prototype is then measured inside the sphere (see Figure \ref{sphere}) using a microphone located at the sphere's center. The main objective of this measurement is to assess the relevance, realism, and performance of the 3D sound synthesis, considering the room and experimental conditions.
Figure \ref{result} shows the transfer function measured at the sphere's center, resulting from the 3D sound synthesis, compared with a computed transfer function in free field.

A good fit between measurement and free field model is observed from 200 Hz to 10 kHz. However, there are some discrepancies between the two curves, which can be explained by room effects and experimental conditions. Below 200 Hz, the measurement is mainly composed of background noise within the room, since the frequency response of the loudspeakers is limited below 100 Hz. Above 2 kHz, the scatterings due to the spatialization sphere structure and to residual reflections in the acoustically treated room are responsible for mismatches between measurement and synthesis.

Those room effects and phenomena induced by experimental conditions, such as scattering, were quantified by measuring the transfer function of a monopole of amplitude 1 at all frequencies, synthesized by the same spatialization sphere using ambisonics. As shown in Figure 6, they are then very well compensated for the horn by normalizing its measured transfer function by the measured transfer function of the monopole.
\begin{figure}[H]
	\centering
	\includegraphics[width=.67\textwidth]{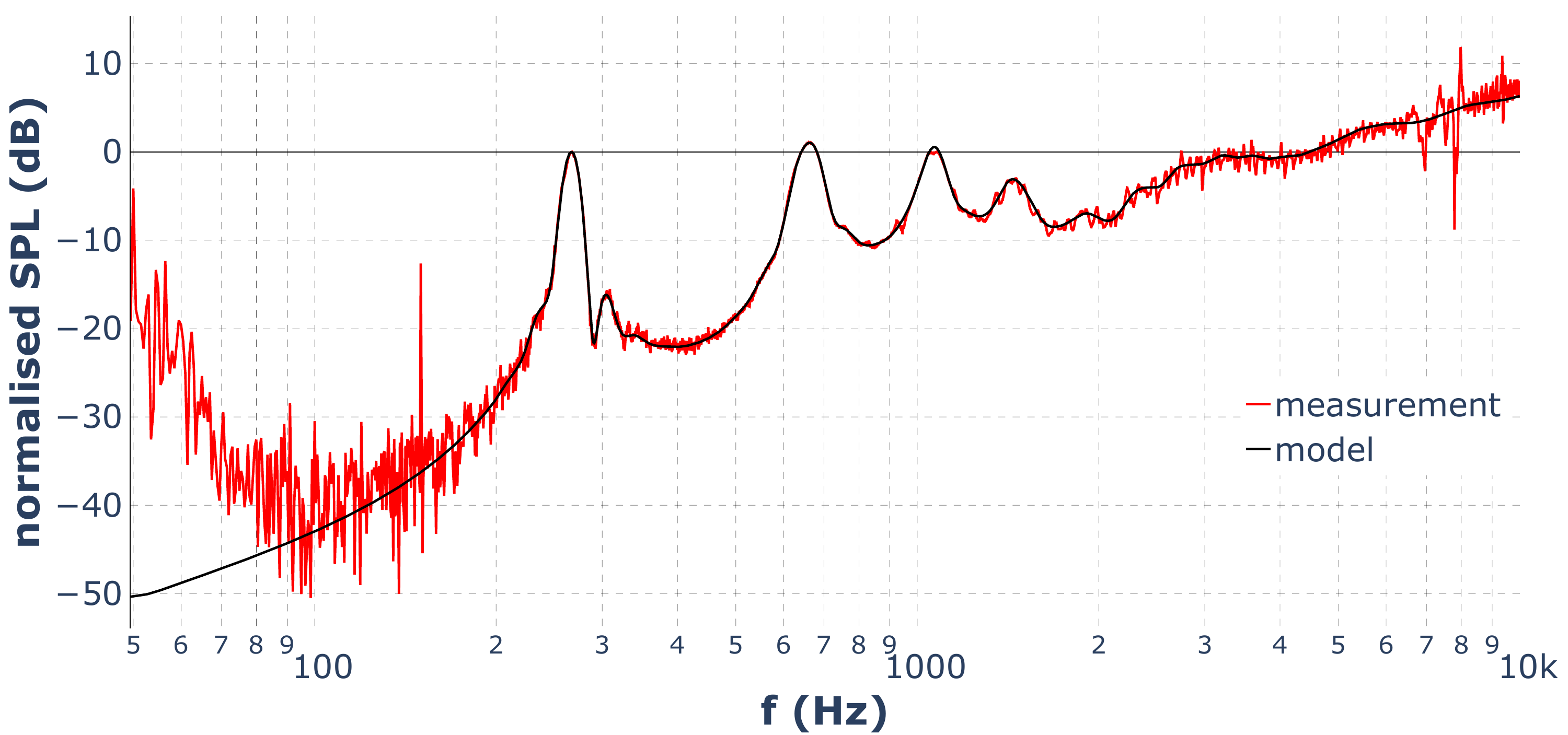}
	\caption{Normalized on-axis Transfer Function of a Bessel horn at 2 m ($r_{in}=2$ cm, $r_{out}=16$ cm, $L=40$ cm)}
	\label{result2}
\end{figure}

%The effects induced by the sphere itself and by the room can be reduced by normalising the measure by the transfer function of a monopole, measured in the same experimental conditions. The improvement brought by this solution is shown in the figure \ref{norm}.
%\begin{figure}[H]
%	\centering
%	\subfloat[without normalisation]{
%		\includegraphics[width=.7\textwidth]{modeleVSmesure2} 	\label{nonorm}
%	} \\
%	\subfloat[with normalisation by the monopole measurements]{
%		\includegraphics[width=.7\textwidth]{modeleVSmesure} \label{norm}
%	}
%	\caption{On-axis Transfer Function at 2 m of a Bessel horn ($r_{in}=2$ cm, $r_{out}=16$ cm, $L=40$ cm)}
%	\label{result}
%\end{figure}

%==============================================================================
\section{Conclusion}
The performance of the 3D soundfield synthesis was assessed using the multimodal formulation and the higher-order ambisonic technique. The pressure inside the horn was computed using the general multimodal formulation developed for a variable section waveguide, and the radiated pressure field was simulated using the multimodal method with a Perfectly Matched Layer (PML).

The proposed methodology in this paper yields realistic and relevant results since the synthesis fits the measurement within the frequency bandwidth of the loudspeakers used for the soundfield reproduction. Discrepancies have been identified below 100 Hz due to the inherent characteristics of loudspeakers in this frequency range and the background noise of the room. Other deviations can be observed above 2 kHz due to scattering by the sphere's structure and residual room effects.

By normalizing the virtual horn measurement by the frequency response of a monopole spatialized in the same conditions, the discrepancies above 2 kHz have effectively been reduced.

%==============================================================================
\section{Future works}
Dong et al. \cite{dong2024} has recently developed a formalism based on the calculation of the Helmholtz-Kirchoff integral from the pressures and gradients calculated numerically around the horn using the multimodal method. In future developments associated with the research presented in this article, a modulus and phase directivity function will be implemented using the same methodology.

We are also planning to model several sources placed above a reflecting ground and encode them in the ambisonic domain in order to synthesize three-dimensional outdoor sound scenes.

%==============================================================================
%\newpage
\bibliographystyle{unsrt}
\bibliography{refs} 

\end{document}